\documentclass[11pt]{article}
\usepackage{amssymb}
\usepackage{amsmath}

\begin{document}

\title{ Supersymmetric Electroweak Corrections to the Higgs Boson Decays into
       Chargino or Neutralino Pair
\footnote{Supported by National Natural Science Foundation of China.}}
\vspace{3mm}

\author{{ Zhang Ren-You$^{2}$, Ma Wen-Gan$^{1,2}$, Wan Lang-Hui$^{2}$, and Jiang Yi$^{2}$ }\\
{\small $^{1}$ CCAST (World Laboratory), P.O.Box 8730, Beijing 100080, P.R.China} \\
{\small $^{2}$ Department of Modern Physics, University of Science and Technology}\\
{\small of China (USTC), Hefei, Anhui 230027, P.R.China} }

\date{}
\maketitle
\vskip 12mm

\begin{abstract}
We investigate the supersymmetric electroweak corrections to the
decay widths of the $CP$-odd and the heavy $CP$-even Higgs bosons
into chargino or neutralino pair in the framework of the Minimal
Supersymmetric Standard Model. The corrections involve the
contributions of the order $O(\alpha_{ew} m_{t(b)}^3/m_W^3)$,
$O(\alpha_{ew} m_{t(b)}^2/m_W^2)$ and $O(\alpha_{ew}
m_{t(b)}/m_W)$. The detailed calculations of the electroweak
corrections to the following decay processes: $A^0/H^0 \rightarrow
\tilde{\chi}^+_1 \tilde{\chi}^-_1$ and $A^0/H^0 \rightarrow
\tilde{\chi}^0_2 \tilde{\chi}^0_2$ are presented in this paper. We
find that these relative corrections maybe rather large
quantitatively, and can exceed $10 \%$ in some regions of
parameter space. The corrections to the decay $A^0/H^0 \rightarrow
\tilde{\chi}^0_1 \tilde{\chi}^0_2$ can be obtained analogously,
but our results show that they are very small and can be
neglected.
\end{abstract}

\vskip 5cm

{\large\bf PACS: 12.15.Lk, 12.60.Jv, 14.80.Cp, 14.80.Ly}

\vfill \eject

\baselineskip=0.36in

\renewcommand{\theequation}{\arabic{section}.\arabic{equation}}
\newcommand{\nb}{\nonumber}

\makeatletter      
\@addtoreset{equation}{section}
\makeatother       

\section{Introduction}
\par
In the Standard Model (SM) \cite{sm}, there exist quadratically
divergent contributions to the corrections to the Higgs boson
mass. This is the so-called naturalness problem of the SM. A good
way to solve this problem is to consider supersymmetric (SUSY)
extensions of the SM. Then the quadratic divergences can be
cancelled by loop diagrams involving the supersymmetric partners
of the SM particles exactly. The most attractive supersymmetric
extension of the SM is the Minimal Supersymmetric Standard Model
(MSSM) \cite{mssm-1,mssm-2}. In the MSSM, we need two Higgs
doublets $H_1$ and $H_2$ to give masses to up and down type
fermions and to assure the cancellation of anomalies. The Higgs
spectrum consists of three neutral Higgs bosons: one $CP$-odd
particle ($A^0$), two $CP$-even particles ($h^0$ and $H^0$), and a
charged Higgs boson ($H^{\pm}$). At the tree level, the neutral
Higgs boson masses are determined by two input parameters (see
Eq.(\ref{relation of the mass parameters})). In this paper, we
choose the mass of the $CP$-odd Higgs boson $m_A$ and $\tan\beta$
as the two input parameters in Higgs sector, which is defined by
$\tan\beta = v_2/v_1$, where $v_1$ and $v_2$ are the vacuum
expectation values of the two Higgs boson fields $H_1$ and $H_2$,
respectively. The radiative corrections to the Higgs boson masses
have been discussed in Ref.\cite{Mh correc,Dabelstein,SHein}. We
find that the magnitude of the corrections to $m_H$ and $m_A$ are
less than $1 \%$ with our parameter choice, therefore they will
not be considered in our calculation.
\par
The neutral Higgs boson decays can be divided into two types of
decay modes, i.e. the SM decay modes and the SUSY decay modes. The
first type of decay modes (e.g. $A^0/H^0 \rightarrow \tau^+ \tau^-$,
$H^0 \rightarrow \gamma \gamma$ and $ZZ$ or $ZZ^* \rightarrow 4l$)
have been already carefully discussed, especially in the parameter
space with large SUSY particle masses where the Higgs boson decays
into sparticles are kinematically forbidden \cite{smmode1,smmode2}.
But for the second type of decay modes, it has
been found that the decays of the MSSM Higgs bosons into charginos
or neutralinos may have substantial branching fractions when
kinematically allowed, and can sometimes even dominate
the usual SM decay modes $\cite{Bisset-1,Djouadi}$. Moreover,
the one-loop calculation of the Higgs decay into the lightest neutralino
pair $(\tilde{\chi}_1^0 \tilde{\chi}_1^0)$ is important for the dark matter 
search\cite{dark matter1,dark matter2}. Therefore, the
widths of these SUSY decay modes should be calculated more precisely.
In this paper, we will study the radiative corrections to the decays
of $A^0$ and $H^0$ into charginos or neutralinos in detail.
\par
Considering the Yukawa corrections of the order
$O(\alpha_{ew}m_{t(b)}^3/m_W^3)$, $O(\alpha_{ew}m_{t(b)}^2/m_W^2)$
and $O(\alpha_{ew}m_{t(b)}/m_W)$, we need calculate the one-loop
Feynman diagrams involving the third generation of quarks and
squarks which are the main contributions to the electroweak
radiative corrections. The structure of this paper is as follows:
in Section 2, we review briefly chargino-neutralino, squark and
Higgs sectors of the MSSM, and give some results at the
tree-level, in Section 3, we present our renormalization
procedure, especially for the Higgs sector, and give some
analytical results of the corrections to the decay widths, the
numerical results and conclusions are presented in section 4, and
in the Appendix, we present the explicit expressions of form
factors for these decays and some necessary formulas.

\section{Related theories at tree-level}
\subsection{Chargino and Neutralino sector}
The tree-level chargino and neutralino mass matrices $X$, $Y$ are \cite{mssm-2}
\begin{eqnarray}
\label{mxn}
X &=& \left(
          \begin{array}{cc}
                M & \sqrt{2}m_W \sin\beta\\
                \sqrt{2}m_W \cos\beta & \mu
          \end{array}
    \right) \nb \\
Y &=& \left(
          \begin{array}{cccc}
               M^{\prime} & 0 & -m_Z s_W \cos\beta & m_Z s_W \sin\beta  \\
               0 & M & m_Z c_W \cos\beta & -m_Z c_W \sin\beta \\
               -m_Z s_W \cos\beta & m_Z c_W \cos\beta & 0 & -\mu \\
               m_Z s_W \sin\beta & -m_Z c_W \sin\beta & -\mu & 0
          \end{array}
      \right)
\end{eqnarray}
To obtain the tree-level masses of chargino and neutralino, we introduce
three unitary matrices $U, V$ and $N$ to diagonalize $X$ and $Y$:
\begin{eqnarray}
U^*XV^{-1} &=& M_D ={\rm diag}(m_{\tilde{\chi}_1^+}, m_{\tilde{\chi}_2^+}),
~~~~~~~(0<m_{\tilde{\chi}_1^+}<m_{\tilde{\chi}_2^+}) \nb \\
N^*YN^{-1} &=& N_D = {\rm diag}(m_{\tilde{\chi}_1^0}, m_{\tilde{\chi}_2^0},
                                  m_{\tilde{\chi}_3^0}, m_{\tilde{\chi}_4^0}),
(0<m_{\tilde{\chi}_1^0}<m_{\tilde{\chi}_2^0}<m_{\tilde{\chi}_3^0}<
  m_{\tilde{\chi}_4^0}).
\end{eqnarray}
where $m_{\tilde{\chi}^+_i}, (i = 1, 2)$ and $m_{\tilde{\chi}^0_j}, (j = 1,...,4)$
are the masses of charginos and neutralinos respectively.
\subsection{Squark sector}
The tree-level stop and sbottom mass matrices are
\begin{eqnarray}
{\cal M}_{\tilde{t}}^2 &=&
          \left(
          \begin{array}{cc}
          M_{\tilde{Q}}^2+m_t^2+m_Z^2\cos 2\beta \left(\frac{1}{2}-
          \frac{2}{3}s_W^2\right) & m_t \left(A_t-\mu \cot \beta\right) \\
          m_t \left(A_t - \mu \cot \beta\right) & M_{\tilde{U}}^2+m_t^2+
          \frac{2}{3}m_Z^2\cos 2\beta s_W^2
          \end{array}
      \right) \nb \\
{\cal M}_{\tilde{b}}^2 &=& \left( \begin{array}{cc}  M_{\tilde{Q}}^2
+ m_b^2 - m_Z^2 \cos 2\beta \left(\frac{1}{2} - \frac{1}{3}
s_W^2\right) & m_b\left(A_b - \mu \tan \beta\right) \\ m_b
\left(A_b - \mu \tan \beta\right) &
M_{\tilde{D}}^2+m_b^2-\frac{1}{3}m_Z^2\cos 2\beta s_W^2
\end{array}\right)
\end{eqnarray}
where $M_{\tilde{Q}},M_{\tilde{U}},M_{\tilde{D}}$ are the
soft-SUSY-breaking masses and $A_{t,b}$ the
corresponding soft-SUSY-breaking trilinear coupling parameters.
They can be diagonalized by unitary matrices ${\cal R}^{\tilde{q}}$ via
\begin{equation}
{\cal R}^{\tilde{q}} {\cal
M}_{\tilde{q}}^2 {\cal R}^{\tilde{q}\dag} = {\rm diag}(m_{\tilde{q}_1}^2,
m_{\tilde{q}_2}^2)
\end{equation}
where
\begin{equation}
{\cal R}^{\tilde{q}} =
     \left(
          \begin{array}{cc}
          \cos \theta_{q} & \sin \theta_{q} \\
     -\sin \theta_{q} & \cos \theta_{q}
          \end{array}
     \right)
     ~~~~~~(-\pi/2 \le \theta_{\tilde{q}}\le \pi/2,~~~~~ q = t, b).
\end{equation}
\subsection{Higgs sector}
For the convenience of the Higgs potential, we introduce two $SU(2)$ Higgs
doublets $H_L, H_R$ defined as
\begin{eqnarray}
H_2 = H_L, ~~~~~~~~~~ H_1 = H_R^{\rm{c}} = i \tau_2 H_R^*
\end{eqnarray}
where $\tau_2$ is the second Pauli matrix.
The Higgs potential can be divided into four parts
\begin{eqnarray}
\label{Higgs potential}
V_H = V_H^{(1)} + V_H^{(2)} + V_H^{(3)} + V_H^{(4)}
\end{eqnarray}
which represent the linear, quadratic, cube and quartic
terms respectively.
In terms of the shifted fields by \cite{Dabelstein} \cite{smmode2}
\begin{eqnarray}
H_L = \left(
            \begin{array}{c}
            H_{L+} \\
            v_L/\sqrt{2} + H_{L0}^R + i H_{L0}^I
            \end{array}
      \right),~~~~~~~
H_R = \left(
            \begin{array}{c}
            H_{R+} \\
            v_R/\sqrt{2} + H_{R0}^R + i H_{R0}^I
            \end{array}
      \right)
\end{eqnarray}
the linear term can be expressed as
\begin{eqnarray}
\label{Higgs linear term}
V_H^{(1)} &=& \sqrt{2} v_L H_{L0}^R \left[
               m_{LH}^2 + \mid \mu \mid^2 + \frac{g^2 + g^{\prime 2}}{8}
              (v_L^2 - v_R^2) - m^2 \frac{v_R}{v_L} \right] \nb \\
          &+& \sqrt{2} v_R H_{R0}^R \left[
           m_{RH}^2 + \mid \mu \mid^2 + \frac{g^2 + g^{\prime 2}}{8}
              (v_R^2 - v_L^2) - m^2 \frac{v_L}{v_R} \right]
\end{eqnarray}
where $m_{LH}, m_{RH}$ and $m$ are the soft-SUSY-breaking Higgs sector
mass parameters.
We define the coefficients of $V_{L0}^R$ and $V_{R0}^R$ in
Eq.(\ref{Higgs linear term}) to be
\begin{eqnarray}
T_L &=& v_L \left[ m_{LH}^2 + \mid \mu \mid^2 + \frac{g^2 + g^{\prime 2}}{8}
                   (v_L^2 - v_R^2) - m^2 \frac{v_R}{v_L} \right] \nb \\
T_R &=& v_R \left[ m_{RH}^2 + \mid \mu \mid^2 + \frac{g^2 + g^{\prime 2}}{8}
                   (v_R^2 - v_L^2) - m^2 \frac{v_L}{v_R} \right]
\end{eqnarray}
Then the linear term of Higgs potential can be written as
\begin{equation}
\label{VH1}
V_H^{(1)} = \sqrt{2} H_{L0}^R T_L + \sqrt{2} H_{R0}^R T_R.
\end{equation}
\par
The quadratic Higgs potential $V_H^{(2)}$ in Eq.(\ref{Higgs
potential}) consists of two parts which are relevant to charged
Higgs and neutral Higgs bosons, respectively. In this paper, we
only consider the radiative corrections to the MSSM neutral Higgs
boson decays. The neutral Higgs part of $V_H^{(2)}$ can be written
in the form \cite{smmode2}
\begin{eqnarray}
&&V_H^{(2)\rm{neu}} = \left(
                      \begin{array}{cc}
                            H_{L0}^I  H_{R0}^I
                      \end{array}
                 \right)
                 \left(
              \begin{array}{cc}
                    \frac{T_L}{v_L} + m^2 \frac{v_R}{v_L} & -m^2 \\
                            -m^2 & \frac{T_R}{v_R} + m^2 \frac{v_L}{v_R}
              \end{array}
                 \right)
         \left(
              \begin{array}{c}
              H_{L0}^I \\
              H_{R0}^I
              \end{array}
                 \right) \nb \\
             &&+ \left(
                      \begin{array}{cc}
                            H_{L0}^R  H_{R0}^R
                      \end{array}
                 \right)
         \left(
              \begin{array}{cc}
                    \frac{T_L}{v_L} + m^2 \frac{v_R}{v_L} +
                \frac{g^2 + g^{\prime 2}}{4} v_L^2 &
                -m^2 - \frac{g^2 + g^{\prime 2}}{4} v_L v_R \\
                            -m^2 - \frac{g^2 + g^{\prime 2}}{4} v_L v_R &
                \frac{T_R}{v_R} + m^2 \frac{v_L}{v_R} +
                \frac{g^2 + g^{\prime 2}}{4} v_R^2
              \end{array}
                 \right)
         \left(
              \begin{array}{c}
              H_{L0}^R \\
              H_{R0}^R
              \end{array}
                 \right). \nb
\end{eqnarray}
By defining
\begin{eqnarray}
\sqrt{2} \left(
         \begin{array}{c}
     H_{L0}^I \\
     H_{R0}^I
         \end{array}
     \right) &=& S(\beta) \left(
                        \begin{array}{c}
                        G^0 \\
                -A^0
                        \end{array}
                \right)
                 = \left(
           \begin{array}{cc}
           \sin\beta & -\cos\beta \\
           \cos\beta & \sin\beta
           \end{array}
           \right) \left(
                   \begin{array}{c}
                   G^0 \\
               -A^0
                   \end{array}
               \right) \nb \\
\sqrt{2} \left(
         \begin{array}{c}
     H_{L0}^R \\
     H_{R0}^R
         \end{array}
     \right) &=& S(\alpha) \left(
                        \begin{array}{c}
                        H^0 \\
                -h^0
                        \end{array}
                \right)
                 = \left(
           \begin{array}{cc}
           \sin\alpha & -\cos\alpha \\
           \cos\alpha & \sin\alpha
           \end{array}
           \right) \left(
                   \begin{array}{c}
                   H^0 \\
               -h^0
                   \end{array}
               \right)
\end{eqnarray}
\begin{eqnarray}
\label{relation of the mass parameters}
&& \tan\beta = \frac{v_L}{v_R}, ~~~~~~~~~~~~~~~~~
\tan2\alpha = \frac{m_A^2 + m_Z^2}{m_A^2 - m_Z^2} \tan2\beta \nb \\
&& m_A^2 = m^2 \frac{v_L^2 + v_R^2}{v_L v_R} = \frac{m^2}{\sin\beta \cos\beta} \nb \\
&& m_{H,h}^2 = \frac{1}{2} \left(
               m_A^2 + m_Z^2 \pm \sqrt{(m_A^2 + m_Z^2)^2 - 4 m_Z^2 m_A^2 \cos^2(2 \beta)} \right)
\end{eqnarray}
\begin{eqnarray}
T_H &=& T_L \sin\alpha + T_R \cos\alpha \nb \\
T_h &=& T_L \cos\alpha - T_R \sin\alpha,
\end{eqnarray}
we can get
\begin{eqnarray}
\label{V_H^1 and V_H^2neu}
V_H^{(1)} &=& T_H H^0 + T_h h^0 \nb \\
V_H^{(2)\rm{neu}} &=& \frac{1}{2}
        \left(
        \begin{array}{cc}
        H^0 h^0
        \end{array}
        \right)
    \left(
        \begin{array}{cc}
        m_H^2 + b_{HH} & b_{Hh} \\
    b_{Hh} & m_h^2 + b_{hh}
        \end{array}
    \right)
    \left(
    \begin{array}{c}
    H^0 \\
    h^0
    \end{array}
    \right) \nb \\
&+& \frac{1}{2}
        \left(
        \begin{array}{cc}
        G^0 A^0
        \end{array}
        \right)
    \left(
        \begin{array}{cc}
        b_{GG} & b_{GA} \\
    b_{GA} & m_A^2 + b_{AA}
        \end{array}
    \right)
    \left(
    \begin{array}{c}
    G^0 \\
    A^0
    \end{array}
    \right)
\end{eqnarray}
with
\begin{eqnarray}
\label{b function}
b_{HH} &=& \frac{g}{m_W \sin2\beta} [ T_H (\cos^3\alpha \sin\beta + \sin^3\alpha \cos\beta)
                                    + T_h \sin\alpha \cos\alpha \sin(\alpha - \beta) ] \nb \\
b_{hh} &=& \frac{g}{m_W \sin2\beta} [ T_H \sin\alpha \cos\alpha \cos(\alpha - \beta)
                                    + T_h (\cos^3\alpha \cos\beta - \sin^3\alpha \sin\beta) ] \nb \\
b_{Hh} &=& \frac{g \sin2\alpha}{2 m_W \sin2\beta} [ T_H \sin(\alpha - \beta)
                                                  + T_h \cos(\alpha - \beta) ] \nb \\
b_{GG} &=& \frac{g}{2 m_W} [ T_H \cos(\alpha - \beta)
                           - T_h \sin(\alpha - \beta) ] \nb \\
b_{AA} &=& \frac{g}{m_W \sin2\beta} [ T_H (\sin^3\beta \cos\alpha + \cos^3\beta \sin\alpha)
                                    + T_h (\cos^3\beta \cos\alpha - \sin^3\beta \sin\alpha) ] \nb \\
b_{GA} &=& \frac{g}{2 m_W} [ T_H \sin(\alpha - \beta)
                           + T_h \cos(\alpha - \beta) ]
\end{eqnarray}

\subsection{Tree-level results}
\par
The relevant part of the MSSM Lagrangian at the tree-level to the
decays concerned in this paper, can be expressed as
\begin{eqnarray}
\label{relevant L}
{\cal L} &=&  \sum_{i,j=1}^2 [
             - g H^0 \bar{\tilde{\chi}}^+_i(P_L J_{j i}^{H*} + P_R J_{i j}^H)\tilde{\chi}^+_j
             - g h^0 \bar{\tilde{\chi}}^+_i(P_L J_{j i}^{h*} + P_R J_{i j}^h)\tilde{\chi}^+_j \nb \\
        &+& i g G^0 \bar{\tilde{\chi}}^+_i(P_L J_{j i}^{G*} - P_R J_{i j}^G)\tilde{\chi}^+_j
      + i g A^0 \bar{\tilde{\chi}}^+_i(P_L J_{j i}^{A*} - P_R J_{i j}^A)\tilde{\chi}^+_j
                        ] \nb \\
        &+& \frac{1}{2} \sum_{i,j=1}^4 [
            g H^0 \bar{\tilde{\chi}}^0_i(P_L F_{i j}^{H*} + P_R F_{i j}^H)\tilde{\chi}^0_j
      + g h^0 \bar{\tilde{\chi}}^0_i(P_L F_{i j}^{h*} + P_R F_{i j}^h)\tilde{\chi}^0_j \nb \\
        &-& i g G^0 \bar{\tilde{\chi}}^0_i(P_L F_{i j}^{G*} - P_R F_{i j}^G)\tilde{\chi}^0_j
      - i g A^0 \bar{\tilde{\chi}}^0_i(P_L F_{i j}^{A*} - P_R F_{i j}^A)\tilde{\chi}^0_j
                        ]
\end{eqnarray}
where $P_{L,R} = \frac{1}{2}(1 \mp \gamma_5)$,
and the matrices $J_{ij}^{H,h,G,A}, F_{ij}^{H,h,G,A}$ are defined in Appendix A.
\par
By using above Lagrangian, we can easily obtain
\begin{eqnarray}
\label{tree-level amplitude}
\mid {\cal M}_t(A^0 \rightarrow \tilde{\chi}^+_i \tilde{\chi}^-_j) \mid^2
       &=& g^2 \left[ (\mid J_{i j}^A \mid^2 + \mid J_{j i}^A \mid^2) m_A^2
         - (\mid J_{i j}^A m_{\tilde{\chi}^+_i} - J_{j i}^{A*} m_{\tilde{\chi}^+_j} \mid^2
        \right. \nb \\
       &&   \left.
       +\mid J_{i j}^A m_{\tilde{\chi}^+_j} - J_{j i}^{A*} m_{\tilde{\chi}^+_i}\mid^2) \right] \nb \\
\mid {\cal M}_t(H^0 \rightarrow \tilde{\chi}^+_i \tilde{\chi}^-_j) \mid^2
       &=& g^2 \left[ (\mid J_{i j}^H \mid^2 + \mid J_{j i}^H \mid^2) m_H^2
         - (\mid J_{i j}^H m_{\tilde{\chi}^+_i} + J_{j i}^{H*} m_{\tilde{\chi}^+_j} \mid^2
        \right. \nb \\
       &&   \left.
       +\mid J_{i j}^H m_{\tilde{\chi}^+_j} + J_{j i}^{H*} m_{\tilde{\chi}^+_i}\mid^2) \right] \nb \\
\mid {\cal M}_t(A^0 \rightarrow \tilde{\chi}^0_i \tilde{\chi}^0_j) \mid^2
       &=& 2 g^2 \left[ \mid F_{i j}^A \mid^2 m_A^2
         - \mid F_{i j}^A m_{\tilde{\chi}^0_i} - F_{i j}^{A*} m_{\tilde{\chi}^0_j} \mid^2 \right] \nb \\
\mid {\cal M}_t(H^0 \rightarrow \tilde{\chi}^0_i \tilde{\chi}^0_j) \mid^2
       &=& 2 g^2 \left[ \mid F_{i j}^H \mid^2 m_H^2
         - \mid F_{i j}^H m_{\tilde{\chi}^0_i} + F_{i j}^{H*} m_{\tilde{\chi}^0_j} \mid^2 \right]
\end{eqnarray}
The tree-level decay widths can be obtained immediately after the
integration over the phase space.

\section{Renormalization and radiative corrections}
The Feynman diagrams involving the third generation of quarks and
squarks are shown in Fig.1 (Fig.1.c and Fig.1.f). In our
calculation, we use the t'Hooft gauge and adopt the dimensional
reduction (DR) \cite{DR scheme} scheme, which preserves
supersymmetry at least at the one-loop level. The complete
on-mass-shell scheme \cite{on-mass-shell-1,on-mass-shell-2,
on-mass-shell-3} is used in doing renormalization.
\par
The relevant renormalization constants in our calculation are defined as \cite{on-mass-shell-1}:
\begin{eqnarray}
\label{renor const}
P_L \tilde{\chi}_i^+ & \rightarrow & Z^{L1/2}_{+,i j}P_L \tilde{\chi}_j^+,~~~~
P_R \tilde{\chi}_i^+  \rightarrow  Z^{R1/2}_{+,i j}P_R \tilde{\chi}_j^+ \nb \\
P_L \tilde{\chi}_i^0 & \rightarrow & Z^{L1/2}_{0,i j}P_L \tilde{\chi}_j^0,~~~~~
P_R \tilde{\chi}_i^0  \rightarrow  Z^{R1/2}_{0,i j}P_R \tilde{\chi}_j^0 ~~~~~~~~~
(Z^R_{0,i j} = Z^{L*}_{0,i j}) \nb \\
\left(
     \begin{array}{c}
     G^0 \\
     A^0
     \end{array}
\right) & \rightarrow &
\left(
     \begin{array}{cc}
     1 + \frac{1}{2} \delta Z_{GG} & \frac{1}{2} \delta Z_{GA} \\
     \frac{1}{2} \delta Z_{AG} & 1 + \frac{1}{2} \delta Z_{AA}
     \end{array}
\right)
\left(
     \begin{array}{c}
     G^0 \\
     A^0
     \end{array}
\right) \nb \\
\left(
     \begin{array}{c}
     H^0 \\
     h^0
     \end{array}
\right) & \rightarrow &
\left(
     \begin{array}{cc}
     1 + \frac{1}{2} \delta Z_{HH} & \frac{1}{2} \delta Z_{Hh} \\
     \frac{1}{2} \delta Z_{hH} & 1 + \frac{1}{2} \delta Z_{hh}
     \end{array}
\right)
\left(
     \begin{array}{c}
     H^0 \\
     h^0
     \end{array}
\right) \nb \\
g & \rightarrow & g + \delta g, ~~~~~~~~~~~~~~~
\beta \rightarrow \beta + \delta \beta \nb \\
m_A^2 & \rightarrow & m_A^2 + \delta m_A^2, ~~~~~~~~
m_{Z,W}^2 \rightarrow m_{Z,W}^2 + \delta m_{Z,W}^2 \nb \\
U & \rightarrow & U+\delta U,~~~V \rightarrow
V+\delta V,~~~ N \rightarrow N+\delta N.
\end{eqnarray}
All these renormalization constants can be fixed by the on-mass-shell
renormalization conditions:
\par
(1) The vacuum expectation values (VEV) of $H^0$ and $h^0$ are zero \cite{Dabelstein}:
\begin{equation}
\label{vacuum condition}
\langle 0 \mid H^0 \mid 0 \rangle = \langle 0 \mid h^0 \mid 0 \rangle = 0
\end{equation}
At the tree-level, this means $T_H = T_h = 0$, and therefore $b_{HH,Hh,hh,GG,GA,AA} = 0$.
Then from Eq.(\ref{V_H^1 and V_H^2neu}), we can get the counterterm of $V_H^{(1)}$
\begin{equation}
\label{delta V(1) and (2)neu}
\delta V_H^{(1)} = \delta T_H H^0 + \delta T_h h^0
\end{equation}
When we consider the radiative corrections at one-loop level, the vacuum conditions
(\ref{vacuum condition}) tell us that
\begin{eqnarray}
&& i \hat{\Gamma}^H = i T^H - i \delta T_H = 0 \nb \\
&& i \hat{\Gamma}^h = i T^h - i \delta T_h = 0
\end{eqnarray}
where $i \hat{\Gamma}^{H,h}$ are renormalized one-point Green
functions (renormalized tadpoles) of $H^0$ and $h^0$, and $i
T^{H,h}, -i \delta T_{H,h}$ are the contributions from the
one-loop diagrams (tadpole diagrams) and conterterm, respectively.
\par
(2) On shell conditions for MSSM Higgs bosons \cite{Dabelstein} \cite{smmode2}: \\
Analogously, we can also get the counterterm of
$V_H^{(2)\rm{neu}}$ from Eq.(\ref{V_H^1 and V_H^2neu}).
Furthermore, all the renormalized one-particle-irreducible
two-point Green functions of the MSSM neutral Higgs bosons ($i
\hat{\Gamma}^{HH}, i \hat{\Gamma}^{Hh}...$) can be obtained
directly. The expressions of them are given in Appendix B. In our
renormalization scheme, we choose $m_A, m_Z$ as the input
parameters for Higgs sector and set $m_A, m_Z$ to be the physical
masses of $A^0$ and $Z^0$:
\begin{equation}
m_A = m_A^{\rm{phys.}}~~~~~~~~~~~~~~~~~~~m_Z = m_Z^{\rm{phys.}}
\end{equation}
The relevant on-shell renormalization conditions for Higgs sector are
\begin{eqnarray}
i \hat{\Gamma}^{AA}(m_A^2) = i \hat{\Gamma}^{GA}(m_A^2) = 0
&&~~~~~~~~~
\frac{\partial i \hat{\Gamma}^{AA}(p^2)}{\partial p^2} |_{p^2 = m_A^2} = i \nb \\
i \hat{\Gamma}^{HH}(m_H^{\rm{phys.}2}) = i
\hat{\Gamma}^{Hh}(m_H^{\rm{phys.}2}) = 0 &&~~~~~~~~~
\frac{\partial i \hat{\Gamma}^{HH}(p^2)}{\partial p^2} |_{p^2 =
m_H^{\rm{phys.}2}} = i
\end{eqnarray}
where $m_H^{\rm{phys.}}$ is the physical mass of $H^0$. Then we
obtain
\begin{eqnarray}
&& \delta m_A^2 = \Sigma^{AA}(m_A^2) - \delta b_{AA} \nb \\
&& m_H^{\rm{phys.}2} = m_H^2 + \delta m_H^2 - \Sigma^{HH}(m_H^2) + \delta b_{HH} \nb \\
&& \delta Z_{AA} = - \frac{\partial \Sigma^{AA}(p^2)}{\partial
p^2} \mid_{p^2 = m_A^2} ~~~~~~~~~~
   \delta Z_{HH} = - \frac{\partial \Sigma^{HH}(p^2)}{\partial p^2} \mid_{p^2 = m_H^2} \nb \\
&& \delta Z_{GA} = \frac{2}{m_A^2} [\delta b_{GA} - \Sigma^{GA}(m_A^2)] \nb \\
&& \delta Z_{hH} = \frac{2}{m_H^2 - m_h^2} [\delta b_{Hh} -
\Sigma^{Hh}(m_H^2)]
\end{eqnarray}
with the definition
\begin{equation}
\label{def. of delta m_H} \delta m_H^2 = \frac{\partial
m_H^2}{\partial m_Z^2} \delta m_Z^2
             + \frac{\partial m_H^2}{\partial m_A^2} \delta m_A^2
         + \frac{\partial m_H^2}{\partial \beta} \delta \beta
\end{equation}
\par
(3) For the renormalization of the parameters $\beta$ and $\alpha$, we use the relation
between $\delta \beta$ and $\delta Z_{GA}$
\begin{eqnarray}
\delta Z_{GA} = \sin2\beta \left[ 2 \frac{\delta \tan\beta}{\tan\beta}
              + \frac{\delta v_R}{v_R} - \frac{\delta v_L}{v_L} \right] \nb
\end{eqnarray}
and impose $\delta v_L/v_L = \delta v_R/v_R$, then we have
\begin{eqnarray}
&& \delta \beta = \frac{\delta Z_{GA}}{4} \nb \\
&& \delta \alpha = \sin4\alpha \left[
                   \frac{\delta \beta}{\sin4\beta}
         + \frac{m_A^2 \delta m_Z^2 - m_Z^2 \delta m_A^2}{2 (m_A^4 - m_Z^4)}
         \right]
\end{eqnarray}
The renomalization of the rest parameters and fields can
be found in Ref.\cite{on-mass-shell-1, renor condition-1, renor condition-2, aoki}.
\par
By substituting Eq.(\ref{renor const}) into the Lagrangian (\ref{relevant L}),
we can obtain the counterterms of these vertices (see Eq.(\ref{cv}) in Appendix A).
Then the renormalized amplitudes  can be expressed as
\begin{eqnarray}
{\cal M}_{\rm{tot}}={\cal M}_t + {\cal M}_v + {\cal M}_c
\end{eqnarray}
where ${\cal M}_c$ and ${\cal M}_v$ represent the
counterterm and the vertex corrections respectively.
The divergence in ${\cal M}_v$ can be cancelled by that in ${\cal M}_c$
exactly. It can be checked both analytically and numerically.
The form factors of the renormalized amplitudes
${\cal M}_{\rm{tot}}(A^0/H^0 \rightarrow \tilde{\chi}_i^+ \tilde{\chi}_j^-)$ and
${\cal M}_{\rm{tot}}(A^0/H^0 \rightarrow \tilde{\chi}_i^0 \tilde{\chi}_j^0)$
can be written as
\begin{eqnarray}
&& J_{ij}^{(\rm{tot})A,H} = J_{ij}^{A,H} + J_{ij}^{(v)A,H} + J_{ij}^{(c)A,H} \nb \\
&& F_{ij}^{(\rm{tot})A,H} = F_{ij}^{A,H} + F_{ij}^{(v)A,H} + F_{ij}^{(c)A,H}
\end{eqnarray}
where $J_{ij}^{(v)A,H}$ and $F_{ij}^{(v)A,H}$ are the terms
contributed by the vertex corrections. The expressions of these
form factors are listed in Appendix A. Then we can get the
renormalized decay width $\Gamma_{\rm{tot}}$ directly. To describe
the magnitude of the radiative corrections to the decay width, we
define the relative correction $\delta = (\Gamma_{\rm{tot}} -
\Gamma_{\rm{tree}})/ \Gamma_{\rm{tree}}$. This quantity will be
used in the next section.

\section{Numerical results}
In this section, we present some numerical results for the
quark-squark loop corrections to the chargino and nuetralino
decays of $A^0$ and $H^0$. Since we find the correction to the
decay $A^0/H^0 \rightarrow \tilde{\chi}^0_1 \tilde{\chi}^0_2$ is
very small and can be neglected, we present the results only for
the decays of $A^0/H^0 \rightarrow \tilde{\chi}_1^+
\tilde{\chi}_1^-/ \tilde{\chi}_2^0 \tilde{\chi}_2^0$ here. In our
numerical calculation, the SM parameters are taken to be $m_t =
174.3$ GeV, $m_b = 4.3$ GeV, $m_Z = 91.188$ GeV, $m_W = 80.41$ GeV
and $\alpha_{EW} = 1/128$ \cite{data}. And the other MSSM
parameters are determined as follows:
\par
(1) For the parameters $M_{\tilde{Q},\tilde{U},\tilde{D}}$ and $A_{t,b}$
in sqark mass matrices, we set
$M_{\tilde{Q}} = M_{\tilde{U}} = M_{\tilde{D}} = A_t = A_b = 200$ GeV.
\par
(2) In the calculation of the corrections to the decays
$A^0/H^0 \rightarrow \tilde{\chi}_1^+ \tilde{\chi}_1^-$, we choose
$\tan\beta$ and the masses of the two charginos $m_{\tilde{\chi}_{1,2}^+}$
as the input parameters to determine the relevant parameters in the chargino
sector. Then the parameters $\mu$ and $M$ in the
mass matrix of chargino can be extracted from these input parameters.
(In this paper, we assume $\mu$ is negative.)
\par
(3) When we calculate the corrections to the decays $A^0/H^0
\rightarrow \tilde{\chi}_2^0 \tilde{\chi}_2^0$, we need four
independent parameters to determine the neutralino sector. Usually
these parameters are taken as $\tan\beta, M, M^{\prime}$ and
$\mu$. But in our paper, we impose the relation between $M$ and
$M^{\prime}$: $M^{\prime} = (5/3) \tan^2\theta_W M$ \cite{mssm-2}
and take $\tan\beta, \mu$ and $m_{\tilde{\chi}_2^0}$ as the input
parameters.
\par
In Fig.2 and Fig.3 we plot the relative corrections $\delta$ as
functions of $\tan\beta$ for $A^0$ and $H^0$ decays with $m_A =
250$ GeV and $m_H = 250$ GeV respectively. In the two figures, the
solid curves represent the decays $A^0/H^0 \rightarrow
\tilde{\chi}_1^+ \tilde{\chi}_1^-$ with the input parameters
$m_{\tilde{\chi}_1^+} = 110$ GeV, $m_{\tilde{\chi}_2^+} = 300$
GeV, and the dotted curves represent the decays $A^0/H^0
\rightarrow \tilde{\chi}_2^0 \tilde{\chi}_2^0$ where we take
$m_{\tilde{\chi}_2^0} = 100$ GeV, $\mu = -200$ GeV. For the
chargino decay of $A^0 (H^0)$ (solid curve), the relative
correction $\delta$ decrease from $-7.5 \%$ ($-13.2 \%$) to $-4.6
\%$ ($-5.1 \%$) as the increment of $\tan\beta$ from 2 to about 7,
and then increase to $-14.0 \%$ ($-17.8 \%$) as $\tan\beta$ to 36.
The corrections to the chargino decays of $A^0$ and $H^0$ are very
sensitive to the value of $\tan\beta$ and quite similar to each
other. In contrast to the chargino decays of $A^0$ and $H^0$, the
correction to the neutralino decay of $A^0$: $A^0 \rightarrow
\tilde{\chi}_2^0 \tilde{\chi}_2^0$ (dotted curve in Fig.2) is
insensitive to the value of $\tan\beta$. It varies only from $-5.4
\%$ to $-7.3 \%$ as $\tan\beta$ running from 2 to 36. For the
decay $H^0 \rightarrow \tilde{\chi}_2^0 \tilde{\chi}_2^0$, there
are two peaks at $\tan\beta \sim 1.85$ and 30 which corresponding
to the resonance effects at the positions where $m_{H}(250
\rm{GeV})$$\sim 2 \times m_{\tilde{t}_1}(2 \times 124.65
\rm{GeV})$ and $m_{H}(250 \rm{GeV})$$\sim 2 \times
m_{\tilde{b}_2}(2 \times 124.07 \rm{GeV})$ respectively. The
absence of these resonance effects on the dotted curve in Fig.2
($A^0 \rightarrow \tilde{\chi}_2^0 \tilde{\chi}_2^0$) due to the
fact that the couplings $V_{A^0\tilde{t}_i^{\dag}\tilde{t}_j}$ and
$V_{A^0\tilde{b}_i^{\dag}\tilde{b}_j}$ are zero when $i = j$ (see
Eq.(\ref{the relevant vertices})).
\par
In Fig.4 and Fig.5 we present the relative corrections as
functions of $m_A$ for the chargino and neutralino decays of $A^0$
respectively. Here $\tan\beta$ is set to be 4 (solid curves) and
30 (dotted curves) to make comparison. The values of the relevant
input parameters are given in the figures. The peaks at $m_A
\sim$ 348 GeV on the solid curves $(\tan\beta=4)$ in Fig.4 and
Fig.5 just reflect the resonance effect at the position $m_A
\sim 2 \times m_t$(2 $\times$ 174.3 GeV). As shown in the figures,
the relative corrections are insensitive to $m_A$ and less than
$-7 \%$ for almost all the values in the range of 250(220) GeV to 400
GeV except for the region around the resonance position: $m_A \sim
2 \times m_t$. As mentioned in Appendix C, this top quark
resonance effect are remarkable only when $\tan\beta$ is small for
the appearance of $\cot\beta$ in the coupling of the vertex
$A^0-\bar{t}-t$. Therefore, there are no apparent resonance peaks
at $m_A \sim$ 348 GeV on the dotted curves $(\tan\beta=30)$.
\par
In the case of the chargino decay of $A^0$: $A^0 \rightarrow
\tilde{\chi}_1^+ \tilde{\chi}_1^-$, for $\tan\beta=30,
m_{\tilde{\chi}_1^+} = 110$ GeV and $m_{\tilde{\chi}_2^+} = 300$
GeV (dotted curve in Fig.4), the squark masses are
$m_{\tilde{t}_1} = 180.96$ GeV, $m_{\tilde{t}_2} = 322.28$ GeV,
$m_{\tilde{b}_1} = 241.57$ GeV and $m_{\tilde{b}_2} = 160.71$ GeV.
The first squark resonance peak should appear at $m_A \sim
m_{\tilde{b}_1} + m_{\tilde{b}_2} = 402.28$ GeV which is just
beyond the upper limit of $m_A$ (400 GeV), therefore it can not be
displayed completely in Fig.4. As shown in Fig.4, the relative
correction to this decay varies from $-10.8 \%$ to $-16.3 \%$ as
$m_A$ increasing from 250 GeV to 400 GeV.
\par
For the neutralino decay of $A^0$: $A^0 \rightarrow
\tilde{\chi}_2^0 \tilde{\chi}_2^0$, with the input parameters
$\tan\beta = 30, \mu = - 200$ GeV and $m_{\tilde{\chi}_2^0} = 100$
GeV, the squark masses are $m_{\tilde{t}_1} = 179.65$ GeV,
$m_{\tilde{t}_2} = 323.01$ GeV, $m_{\tilde{b}_1} = 262.28$ GeV and
$m_{\tilde{b}_2} = 124.07$ GeV. The sharp peak on the dotted curve
in Fig.5 is just at $m_A = m_{\tilde{b}_1} + m_{\tilde{b}_2} =
386.35$ GeV. The correction increases slowly from $-6 \%$ to $-8
\%$ as the mass of $A^0$ varying from 220 GeV to 370 GeV, and then
increases rapidly as $m_A$ to 386 GeV for the resonance effect
around this position.
\par
Analogously, the $m_H$ dependence of the corrections to the
chargino and neutralino decays of $H^0$ are depicted in Fig.6 and
Fig.7. From Fig.6 (Fig.7) we can see that for the chargino
(neutralino) decay of $H^0$ there are three peaks on the curves.
The peak on the solid curve corresponds to the resonance effect at
the position $m_H$ = 339.88 (315.14) GeV $\sim 2 \times
m_{\tilde{t}_1}$ = 2 $\times$ 170.27 (158.00) GeV in the case of
$\tan\beta=4$. The other two peaks on the dotted curve are just at
$m_H$ = 321.06 (248.09) GeV $\sim 2 \times m_{\tilde{b}_2}$ = 2
$\times$ 160.71 (124.07) GeV and $m_H$ = 361.05 (359.05) GeV $\sim
2 \times m_{\tilde{t}_1}$ = 2 $\times$ 180.96 (179.65) GeV. They
can also be explained by the resonance effects of the chargino
(neutralino) decay of $H^0$ with $\tan\beta=30$. Except for the
resonance effects, the corrections are about $-5 \% \sim -7 \%$
and insensitive to $m_H$ when $\tan\beta=4$. But, for
$\tan\beta=30$, the corrections can exceed $-10 \%$, especially in
the case of the chargino decay of $H^0$, it can reach about $-17
\%$.
\par
The relative corrections to the chargino decays of $A^0$ and $H^0$
versus the lightest chargino mass $m_{\tilde{\chi}_1^+}$
with $\tan\beta=4$ and 30 are
given in Fig.8. The upper two curves represent the cases of
$\tan\beta=4$ and the lower two, $\tan\beta=30$. The radiative corrections
to these decay modes are all insensitive to $m_{\tilde{\chi}_1^+}$.
In the case of $\tan\beta=30$, the relative corrections are considerable.
They can reach about $-11.7 \%$ and $-15.2 \%$ for the chargino
decays of $A^0$ and $H^0$ respectively.
\par
For the neutralino decays of $A^0$ and $H^0$,
$A^0/H^0 \rightarrow \tilde{\chi}_2^0 \tilde{\chi}_2^0$,
we plot the relative corrections as functions of $m_{\tilde{\chi}_2^0}$
in Fig.9. The curves in this figure are quite different in shape from
those in Fig.8. The corrections are sensitive to $m_{\tilde{\chi}_2^0}$,
especially in the case of large $\tan\beta$.
For $\tan\beta=30$, there are two small peaks on the solid and dotted
curves which are just at $m_{\tilde{\chi}_2^0} = 128$ GeV
$\sim m_b + m_{\tilde{b}_2}$ = 4.3 GeV + 124.07 GeV. They
are the only resonance peaks which can appear in this figure as
$m_{\tilde{\chi}_2^0}$ in the range of 100 to 160 GeV.
\par
In summary, we have computed the supersymmetric electroweak
corrections to the partial widths of the deacys
$A^0/H^0 \rightarrow \tilde{\chi}_1^+ \tilde{\chi}_1^-$ and
$A^0/H^0 \rightarrow \tilde{\chi}_2^0 \tilde{\chi}_2^0$
in the MSSM. Only the third generation of quarks and squarks
are included in our one-loop Feynman diagrams calculation.
The corrections to these decay modes are all important and can
exceed $-10 \%$ in some parameter space, therefore they should
be taken into account in the precise experiment analysis. 
\par
After the submission of this manuscript to Physical Review D, we 
found similar calculations in Ref.\cite{last} where they gave the 
electroweak corrections to the decays $(h^0, H^0, A^0) \rightarrow
\tilde{\chi}^0_i + \tilde{\chi}^0_j$ and 
$\tilde{\chi}^0_i \rightarrow (h^0, H^0, A^0) + \tilde{\chi}^0_j, 
(i,j=1,2,3,4)$.
\par
\noindent{\large\bf Acknowledgements:} This work was supported in
part by the National Natural Science Foundation of China(project
Nos: 19875049, 10005009), the Doctoral Program Foundation of the
Education Ministry of China, and a grant from the Ministry of
Science and Technology of China.

\section{Appendix}
\subsection{Appendix A}
The matrices $J^{H,h,G,A}$ and $F^{H,h,G,A}$ are defined as
follows:
\begin{eqnarray}
&& \left(
     \begin{array}{c}
     J_{i j}^H \\
     J_{i j}^h
     \end{array}
\right) = \left(
     \begin{array}{cc}
     \cos\alpha & \sin\alpha \\
    -\sin\alpha & \cos\alpha
     \end{array}
\right) \left(
     \begin{array}{c}
     Q_{i j} \\
     S_{i j}
     \end{array}
\right) \nb \\
&& \left(
     \begin{array}{c}
     J_{i j}^G \\
     J_{i j}^A
     \end{array}
\right) = \left(
     \begin{array}{cc}
     -\cos\beta & \sin\beta \\
      \sin\beta & \cos\beta
     \end{array}
\right) \left(
     \begin{array}{c}
     Q_{i j} \\
     S_{i j}
     \end{array}
\right) \nb \\
&& \left(
     \begin{array}{c}
     F_{i j}^H \\
     F_{i j}^h
     \end{array}
\right) = \left(
     \begin{array}{cc}
     -\cos\alpha & \sin\alpha \\
      \sin\alpha & \cos\alpha
     \end{array}
\right) \left(
     \begin{array}{c}
     Q_{i j}^{\prime \prime} \\
     S_{i j}^{\prime \prime}
     \end{array}
\right) \nb \\
&& \left(
     \begin{array}{c}
     F_{i j}^G \\
     F_{i j}^A
     \end{array}
\right) = \left(
     \begin{array}{cc}
     \cos\beta & \sin\beta \\
    -\sin\beta & \cos\beta
     \end{array}
\right) \left(
     \begin{array}{c}
     Q_{i j}^{\prime \prime} \\
     S_{i j}^{\prime \prime}
     \end{array}
\right)
\end{eqnarray}
where
\begin{eqnarray}
&&Q_{i j} = \frac{1}{\sqrt{2}} V_{i 1} U_{j 2} \nb \\
&&S_{i j} = \frac{1}{\sqrt{2}} V_{i 2} U_{j 1}
             ~~~~~~~~~~~~~~~~~~~~~~~~~~~~~~~~~~~~~~~~~~(i,j = 1,2) \nb \\
&&Q_{i j}^{\prime \prime} = \frac{1}{2} [ N_{i 3} (N_{j 2} - N_{j
1} \tan\theta_W)
                                        + N_{j 3} (N_{i 2} - N_{i 1} \tan\theta_W) ] \nb \\
&&S_{i j}^{\prime \prime} = \frac{1}{2} [ N_{i 4} (N_{j 2} - N_{j
1} \tan\theta_W)
                                        + N_{j 4} (N_{i 2} - N_{i 1} \tan\theta_W) ]
                    ~~~~(i,j = 1,...,4)
\end{eqnarray}
From these definitions, we can find that $Q^{\prime \prime},
S^{\prime \prime}$ and $F^{H,h,G,A}$ are symmetric matrices. The
counterterms of the vertices
$A^0(H^0)\bar{\tilde{\chi}}_i^+\tilde{\chi}_j^+$ and
$A^0(H^0)\bar{\tilde{\chi}}_i^0\tilde{\chi}_j^0$ can be written in
the form
\begin{eqnarray}
\label{cv} \delta {\cal L} &=& -g H^0 \bar{\tilde{\chi}}_i^+ (P_L
J_{ji}^{(c)H*}
                    +P_R J_{ij}^{(c)H}) \tilde{\chi}_j^+
            + i g A^0 \bar{\tilde{\chi}}_i^+ (P_L J_{ji}^{(c)A*}
            -P_R J_{ij}^{(c)A}) \tilde{\chi}_j^+ \nb \\
                & & +\frac{g}{2} H^0 \bar{\tilde{\chi}}_i^0 (P_L F_{ij}^{(c)H*}
            +P_R F_{ij}^{(c)H}) \tilde{\chi}_j^0
            -\frac{i g}{2} A^0 \bar{\tilde{\chi}}_i^0 (P_L F_{ij}^{(c)A*}
            -P_R F_{ij}^{(c)A}) \tilde{\chi}_j^0
\end{eqnarray}
where $J_{ij}^{(c)A,H}$ and $F_{ij}^{(c)A,H}$ are the form factors
of the counterterms. \\
The expression of $J_{ij}^{(c)H}$ is
\begin{eqnarray}
J_{ij}^{(c)H} = \delta J_{ij}^H + \frac{\delta g}{g} J_{ij}^H
                    +\frac{1}{2} \delta Z_{HH} J_{ij}^H
            +\frac{1}{2} \delta Z_{hH} J_{ij}^h
                    +\sum_{k=1}^2 \left[
            \frac{1}{2} \delta Z_{+,kj}^R J_{ik}^H
            +\frac{1}{2} \delta Z_{+,ki}^{L*} J_{kj}^H \right]
\end{eqnarray}
The other three form factors $J_{ij}^{(c)A}, F_{ij}^{(c)H}$ and
$F_{ij}^{(c)A}$ can be written down analogously. To express the
form factors $F_{ij}^{v(A,H)}$ and $J_{ij}^{v(A,H)}$ contributed
by the vertex corrections, we introduce some notations to denote
the vertices which will be used
\begin{eqnarray}
& & A^0-\bar{t}-t:~~~V_{A^0\bar{t}t}\gamma^5
~~~~~~~~~~~~~~~~~~~~~~~~
    A^0-\bar{b}-b:~~~V_{A^0\bar{b}b}\gamma^5 \nb \\
& & H^0-\bar{t}-t:~~~V_{H^0\bar{t}t} ~~~~~~~~~~~~~~~~~~~~~~~~~~
    H^0-\bar{b}-b:~~~V_{H^0\bar{b}b} \nb \\
& & A^0(H^0)-\tilde{t}_i^{\dag}-\tilde{t}_j:~~~
                          V_{A^0(H^0)\tilde{t}_i^{\dag}\tilde{t}_j}~~~~~~~~~~~
    A^0(H^0)-\tilde{b}_i^{\dag}-\tilde{b}_j:~~~
                          V_{A^0(H^0)\tilde{b}_i^{\dag}\tilde{b}_j} \nb \\
& & \bar{t}-\tilde{\chi}_i^0-\tilde{t}_j:~
                          V_{\bar{t}\tilde{\chi}_i^0\tilde{t}_j}^L P_L
                         +V_{\bar{t}\tilde{\chi}_i^0\tilde{t}_j}^R P_R
~~~~~~~~~ \bar{b}-\tilde{\chi}_i^0-\tilde{b}_j:~
                          V_{\bar{b}\tilde{\chi}_i^0\tilde{b}_j}^L P_L
                         +V_{\bar{b}\tilde{\chi}_i^0\tilde{b}_j}^R P_R \nb \\
& & \bar{t}-\tilde{\chi}_i^+-\tilde{b}_j:~
                          V_{\bar{t}\tilde{\chi}_i^+\tilde{b}_j}^{L} P_L
                         +V_{\bar{t}\tilde{\chi}_i^+\tilde{b}_j}^{R} P_R
~~~~~~~ \bar{b}-\bar{\tilde{\chi}}_i^+-\tilde{t}_j:~
                         (V_{\bar{b}\bar{\tilde{\chi}}_i^+\tilde{t}_j}^L P_L
             +V_{\bar{b}\bar{\tilde{\chi}}_i^+\tilde{t}_j}^R P_R) C~~~~
\end{eqnarray}
For the decays $A^0/H^0 \rightarrow \tilde{\chi}_i^0
\tilde{\chi}_j^0$, we define
\begin{eqnarray}
&& B_0 = B_0(q) \equiv B_0(-p, m_q, m_q) \nb \\
&& C_{0,11,12}^{(1)} = C_{0,11,12}^{(1)}(i,j,q,k) \equiv
   C_{0,11,12}(k_i, -p, m_{\tilde{q}_k}, m_q, m_q) \nb \\
&& C_{0,11,12}^{(2)} = C_{0,11,12}^{(2)}(i,j,q,k,l) \equiv
   C_{0,11,12}(k_i, -p, m_q, m_{\tilde{q}_k}, m_{\tilde{q}_l})
\end{eqnarray}
\begin{eqnarray}
\label{def of F1} F_a^{A,H(1)}=F_a^{A,H(1)}(i,j,q,k) \equiv
V_{A^0,H^0\bar{q}q}
                               V_{\bar{q}\tilde{\chi}_i^0\tilde{q}_k}^{L*}
                               V_{\bar{q}\tilde{\chi}_j^0\tilde{q}_k}^{L} \nb \\
F_b^{A,H(1)}=F_b^{A,H(1)}(i,j,q,k) \equiv V_{A^0,H^0\bar{q}q}
                               V_{\bar{q}\tilde{\chi}_i^0\tilde{q}_k}^{L*}
                               V_{\bar{q}\tilde{\chi}_j^0\tilde{q}_k}^{R} \nb \\
F_c^{A,H(1)}=F_c^{A,H(1)}(i,j,q,k) \equiv V_{A^0,H^0\bar{q}q}
                               V_{\bar{q}\tilde{\chi}_i^0\tilde{q}_k}^{R*}
                               V_{\bar{q}\tilde{\chi}_j^0\tilde{q}_k}^{L} \nb \\
F_d^{A,H(1)}=F_d^{A,H(1)}(i,j,q,k) \equiv V_{A^0,H^0\bar{q}q}
                               V_{\bar{q}\tilde{\chi}_i^0\tilde{q}_k}^{R*}
                               V_{\bar{q}\tilde{\chi}_j^0\tilde{q}_k}^{R}
\end{eqnarray}
\begin{eqnarray}
\label{def of F2} F_a^{A,H(2)}=F_a^{A,H(2)}(i,j,q,k,l) \equiv
                     V_{A^0,H^0\tilde{q}_l^{\dag}\tilde{q}_k}
                     V_{\bar{q}\tilde{\chi}_i^0\tilde{q}_k}^{L*}
                     V_{\bar{q}\tilde{\chi}_j^0\tilde{q}_l}^{L} \nb \\
F_b^{A,H(2)}=F_b^{A,H(2)}(i,j,q,k,l) \equiv
                     V_{A^0,H^0\tilde{q}_l^{\dag}\tilde{q}_k}
                     V_{\bar{q}\tilde{\chi}_i^0\tilde{q}_k}^{L*}
                     V_{\bar{q}\tilde{\chi}_j^0\tilde{q}_l}^{R} \nb \\
F_c^{A,H(2)}=F_c^{A,H(2)}(i,j,q,k,l) \equiv
                     V_{A^0,H^0\tilde{q}_l^{\dag}\tilde{q}_k}
                     V_{\bar{q}\tilde{\chi}_i^0\tilde{q}_k}^{R*}
                     V_{\bar{q}\tilde{\chi}_j^0\tilde{q}_l}^{R}
\end{eqnarray}
where $q=t$ or $b$, $p=k_i+k_j$, $k_i$ and $k_j$ are the outgoing
momenta of $\tilde{\chi}_i^0$ and $\tilde{\chi}_j^0$. Then the
form factors $F_{ij}^{v(A,H)}$ can be written as
\begin{eqnarray}
F_{ij}^{v(A)}= \frac{- N_c}{16 \pi^2 g} \sum_{q=t,b}
\left( \Lambda_{ij}^{A(1)} + \Lambda_{ij}^{A(2)} \right) + (i \leftrightarrow j) \nb \\
F_{ij}^{v(H)}= \frac{-i N_c}{16 \pi^2 g} \sum_{q=t,b} \left(
\Lambda_{ij}^{H(1)} + \Lambda_{ij}^{H(2)} \right) + (i
\leftrightarrow j)
\end{eqnarray}
where $N_c = 3$,
\begin{eqnarray}
\Lambda_{ij}^{A(1)}&=& \sum_{k=1}^2
     \left[
     \left(
     m_q m_{\tilde{\chi}_j^0} F_a^{A(1)}
     +(m_q^2 - m_{\tilde{q}_k}^2) F_b^{A(1)}
     +m_{\tilde{\chi}_i^0} m_{\tilde{\chi}_j^0} F_c^{A(1)}
     +m_q m_{\tilde{\chi}_i^0} F_d^{A(1)}
     \right) C_0^{(1)}
     \right. \nb \\
&&   \left.
     +\left(
     m_{\tilde{\chi}_i^0} m_{\tilde{\chi}_j^0} F_c^{A(1)}
     -m_{\tilde{\chi}_i^0}^2 F_b^{A(1)}
     \right) C_{11}^{(1)}
     +(m_{\tilde{\chi}_i^0}^2 - m_{\tilde{\chi}_j^0}^2) F_b^{A(1)} C_{12}^{(1)}
     +F_b^{A(1)} B_0
     \right] \nb \\
\Lambda_{ij}^{A(2)}&=& \sum_{k,l=1}^2
     \left[
     m_q F_b^{A(2)} C_0^{(2)} - m_{\tilde{\chi}_i^0} F_c^{A(2)} C_{11}^{(2)}
     +\left(
     m_{\tilde{\chi}_i^0} F_c^{A(2)} - m_{\tilde{\chi}_j^0} F_a^{A(2)}
     \right)C_{12}^{(2)}
     \right]
\end{eqnarray}
\begin{eqnarray}
\Lambda_{ij}^{H(1)}&=& \sum_{k=1}^2
     \left[
     \left(
     m_q m_{\tilde{\chi}_j^0} F_a^{H(1)}
     +(m_q^2 + m_{\tilde{q}_k}^2) F_b^{H(1)}
     +m_{\tilde{\chi}_i^0} m_{\tilde{\chi}_j^0} F_c^{H(1)}
     +m_q m_{\tilde{\chi}_i^0} F_d^{H^(1)}
     \right) C_0^{(1)}
     \right. \nb \\
&&   \left.
     +\left(
     m_{\tilde{\chi}_i^0} m_{\tilde{\chi}_j^0} F_c^{H(1)}
     +m_{\tilde{\chi}_i^0}^2 F_b^{H(1)}
     +2 m_q m_{\tilde{\chi}_i^0} F_d^{H(1)}
     \right) C_{11}^{(1)}
     \right. \nb \\
&&   \left.
     +\left(
     2 m_q m_{\tilde{\chi}_j^0} F_a^{H(1)}
     +(m_{\tilde{\chi}_j^0}^2 - m_{\tilde{\chi}_i^0}^2) F_b^{H(1)}
     -2 m_q m_{\tilde{\chi}_i^0} F_d^{H(1)}
     \right) C_{12}^{^(1)}
     -F_b^{H(1)} B_0
     \right] \nb \\
     \Lambda_{ij}^{H(2)}&=& \sum_{k,l=1}^2
     \left[
     m_q F_b^{H(2)} C_0^{(2)}
     -m_{\tilde{\chi}_i^0} F_c^{H(2)} C_{11}^{(2)}
     +\left(
     m_{\tilde{\chi}_i^0} F_c^{H(2)} - m_{\tilde{\chi}_j^0} F_a^{H(2)}
     \right) C_{12}^{(2)}
     \right]
\end{eqnarray}
For the decays $A^0/H^0 \rightarrow \tilde{\chi}_i^+
\tilde{\chi}_j^-$, we define
\begin{eqnarray}
C_{0,11,12}^{\prime (1)}=C_{0,11,12}^{\prime (1)}(i,j,q,k)
          \equiv
      \left\{
      \begin{array} {c}
      C_{0,11,12}(k_i, -p, m_{\tilde{b}_k}, m_t, m_t), ~~~~~q=t \\
          C_{0,11,12}(k_i, -p, m_{\tilde{t}_k}, m_b, m_b), ~~~~~q=b
      \end{array}
      \right. \nb \\
C_{0,11,12}^{\prime (2)}=C_{0,11,12}^{\prime (2)}(i,j,q,k,l)
          \equiv
      \left\{
      \begin{array}{c}
      C_{0,11,12}(k_i, -p, m_t, m_{\tilde{b}_k}, m_{\tilde{b}_l}), ~~~~~q=t \\
          C_{0,11,12}(k_i, -p, m_b, m_{\tilde{t}_k}, m_{\tilde{t}_l}), ~~~~~q=b
      \end{array}
      \right.
\end{eqnarray}
\begin{eqnarray}
J_a^{A,H(1)}=J_a^{A,H(1)}(i, j, q, k) \equiv
          \left\{
      \begin{array}{c}
      V_{A^0,H^0\bar{t}t} V_{\bar{t}\tilde{\chi}_i^+\tilde{b}_k}^{L*}
                      V_{\bar{t}\tilde{\chi}_j^+\tilde{b}_k}^{L},~~~~q=t \\
          V_{A^0,H^0\bar{b}b} V_{\bar{b}\bar{\tilde{\chi}}_i^+\tilde{t}_k}^{L*}
                      V_{\bar{b}\bar{\tilde{\chi}}_j^+\tilde{t}_k}^{L},~~~~q=b
      \end{array}
      \right. \nb \\
J_a^{A,H(2)}=J_a^{A,H(2)}(i, j, q, k, l) \equiv
          \left\{
      \begin{array}{c}
            V_{A^0,H^0\tilde{b}_l^{\dag}\tilde{b}_k}
                V_{\bar{t}\tilde{\chi}_i^+\tilde{b}_k}^{L*}
                V_{\bar{t}\tilde{\chi}_j^+\tilde{b}_l}^{L},~~~~q=t \\
                V_{A^0,H^0\tilde{t}_l^{\dag}\tilde{t}_k}
        V_{\bar{b}\bar{\tilde{\chi}}_i^+\tilde{t}_k}^{L*}
        V_{\bar{b}\bar{\tilde{\chi}}_j^+\tilde{t}_l}^{L},~~~~q=b
      \end{array}
      \right.
\end{eqnarray}
where $k_i$ and $k_j$ are the outgoing momenta of
$\tilde{\chi}_i^+$ and $\tilde{\chi}_j^-$. As in Eq.(\ref{def of
F1}) and (\ref{def of F2}), the definitions of
$J_{b,c,d}^{A,H(1)}$ and $J_{b,c}^{A,H(2)}$ can be obtained from
the definitions of $J_a^{A,H(1)}$ and $J_a^{A,H(2)}$ only by
substituting for their superscripts $L,L$. Then the form factors
$J_{ij}^{v(A,H)}$ can be expressed as follows
\begin{eqnarray}
J_{ij}^{v(A)}=\frac{N_c}{16 \pi^2 g} \left[
         (\Pi_{ij}^{A(1)} + \Pi_{ij}^{A(2)})\mid_{q=t} +
         (\Pi_{ji}^{A(1)} + \Pi_{ji}^{A(2)})\mid_{q=b} \right] \nb \\
J_{ij}^{v(H)}=\frac{i N_c}{16 \pi^2 g} \left[
         (\Pi_{ij}^{H(1)} + \Pi_{ij}^{H(2)})\mid_{q=t} +
         (\Pi_{ji}^{H(1)} + \Pi_{ji}^{H(2)})\mid_{q=b} \right]
\end{eqnarray}
\begin{eqnarray}
\Pi_{ij}^{A(1)} &=& \sum_{k=1}^2
         \left[
     \left(
     m_q m_{\tilde{\chi}_j^+} J_a^{A(1)}
     + \epsilon^A J_b^{A(1)}
     +m_{\tilde{\chi}_i^+} m_{\tilde{\chi}_j^+} J_c^{A(1)}
     +m_q m_{\tilde{\chi}_i^+} J_d^{A(1)}
     \right) C_0^{\prime(1)}
     \right. \nb \\
&&       \left.
     +\left(
     m_{\tilde{\chi}_i^+} m_{\tilde{\chi}_j^+} J_c^{A(1)}
     -m_{\tilde{\chi}_i^+}^2 J_b^{A(1)}
     \right) C_{11}^{\prime(1)}
     +(m_{\tilde{\chi}_i^+}^2 - m_{\tilde{\chi}_j^+}^2) J_b^{A(1)} C_{12}^{\prime(1)}
     +J_b^{A(1)} B_0
     \right] \nb \\
\Pi_{ij}^{A(2)} &=& \sum_{k,l=1}^2
         \left[
     m_q J_b^{A(2)} C_0^{\prime(2)}
     -m_{\tilde{\chi}_i^+} J_c^{A(2)} C_{11}^{\prime(2)}
     +\left(
         m_{\tilde{\chi}_i^+} J_c^{A(2)} - m_{\tilde{\chi}_j^+} J_a^{A(2)}
     \right) C_{12}^{\prime(2)}
     \right]
\end{eqnarray}
\begin{eqnarray}
\Pi_{ij}^{H(1)} &=& \sum_{k=1}^2
         \left[
     \left(
     m_q m_{\tilde{\chi}_j^+} J_a^{H(1)}
     +\epsilon^H J_b^{H(1)}
     +m_{\tilde{\chi}_i^+} m_{\tilde{\chi}_j^+} J_c^{H(1)}
     +m_q m_{\tilde{\chi}_i^+} J_d^{H(1)}
     \right) C_0^{\prime(1)}
     \right. \nb \\
&&       \left.
         +\left(
     m_{\tilde{\chi}_i^+}^2 J_b^{H(1)}
     +m_{\tilde{\chi}_i^+} m_{\tilde{\chi}_j^+} J_c^{H(1)}
     +2 m_q m_{\tilde{\chi}_i^+} J_d^{H(1)}
     \right) C_{11}^{\prime(1)}
     \right. \nb \\
&&       \left.
         +\left(
         2 m_q m_{\tilde{\chi}_j^+} J_a^{H(1)}
     +(m_{\tilde{\chi}_j^+}^2 - m_{\tilde{\chi}_i^+}^2) J_b^{H(1)}
     - 2 m_q m_{\tilde{\chi}_i^+} J_d^{H(1)}
     \right) C_{12}^{\prime(1)}
     -J_b^{H(1)} B_0
     \right] \nb \\
\Pi_{ij}^{H(2)} &=& \sum_{k,l=1}^2
         \left[
     m_q J_b^{H(2)} C_0^{\prime(2)}
     -m_{\tilde{\chi}_i^+} J_c^{H(2)} C_{11}^{\prime(2)}
     +\left(
     m_{\tilde{\chi}_i^+} J_c^{H(2)}
     -m_{\tilde{\chi}_j^+} J_a^{H(2)}
     \right) C_{12}^{\prime(2)}
     \right]
\end{eqnarray}
where
\begin{eqnarray}
\epsilon^{A,H} = \epsilon^{A,H}(q,k) \equiv \left\{
             \begin{array}{c}
             m_t^2 \mp m_{\tilde{b}_k}^2, ~~~~~~q=t \\
         m_b^2 \mp m_{\tilde{t}_k}^2, ~~~~~~q=b
             \end{array}
             \right.
\end{eqnarray}
In this paper, we adopt the definitions and calculation formulas
of the two-point and three-point Passarino-Veltman integrals shown
in Ref.\cite{Bernd} and \cite{Velt}, respectively.

\subsection{Appendix B}
The counterterm of the potential $V_H^{(2)\rm{neu}}$ is
\begin{eqnarray}
\label{delta V_H^2neu} \delta V_H^{(2)\rm{neu}}&=&
    \frac{1}{2} H^{02} (\delta Z_{HH} m_H^2 + \delta m_H^2 + \delta b_{HH})
  + \frac{1}{2} h^{02} (\delta Z_{hh} m_h^2 + \delta m_h^2 + \delta b_{hh}) \nb \\
&+& \frac{1}{2} H^0 h^0 (\delta Z_{Hh} m_H^2 + \delta Z_{hH} m_h^2
+ 2 \delta b_{Hh})
  + \frac{1}{2} G^{02} \delta b_{GG} \nb \\
&+& \frac{1}{2} A^{02} (\delta Z_{AA} m_A^2 + \delta m_A^2 +
\delta b_{AA})
  + \frac{1}{2} G^0 A^0 (\delta Z_{AG} m_A^2 + 2 \delta b_{GA})
\end{eqnarray}
Here $\delta b_{HH}, \delta b_{hh},...$ are functions of $\delta
T_{H}$ and $\delta T_h$. We can get their expressions from that of
$b_{HH}, b_{hh},...$ in Eq.(\ref{b function}) by substituting
$\delta T_H$ and $\delta T_h$ for $T_H$ and $T_h$. Then the
renormalized two-point Green functions can be obtained
\begin{eqnarray}
i \hat{\Gamma}^{HH}(p^2) &=& i (p^2 - m_H^2) + i \Sigma^{HH}(p^2)
     + i [\delta Z_{HH} (p^2 - m_H^2) - \delta m_H^2 - \delta b_{HH}] \nb \\
i \hat{\Gamma}^{hh}(p^2) &=& i (p^2 - m_h^2) + i \Sigma^{hh}(p^2)
     + i [\delta Z_{hh} (p^2 - m_h^2) - \delta m_h^2 - \delta b_{hh}] \nb \\
i \hat{\Gamma}^{Hh}(p^2) &=& i \Sigma^{Hh}(p^2) + i [\frac{1}{2}
\delta Z_{Hh} (p^2 - m_H^2)
                           +\frac{1}{2} \delta Z_{hH} (p^2 - m_h^2)
               -\delta b_{Hh}] \nb \\
i \hat{\Gamma}^{GG}(p^2)
&=& i p^2 + i \Sigma^{GG}(p^2) + i [\delta Z_{GG} p^2 - \delta b_{GG}] \nb \\
i \hat{\Gamma}^{AA}(p^2) &=& i (p^2 - m_A^2) + i \Sigma^{AA}(p^2)
     + i [\delta Z_{AA} (p^2 - m_A^2) - \delta m_A^2 - \delta b_{AA}] \nb \\
i \hat{\Gamma}^{GA}(p^2) &=& i \Sigma^{GA}(p^2) + i [\frac{1}{2}
\delta Z_{GA} p^2 +\frac{1}{2} \delta Z_{AG} (p^2 - m_A^2)
-\delta b_{GA}]
\end{eqnarray}
where the mass parameters of the $CP-$even neutral Higgs bosons $m_H$
and $m_h$ are determined by $m_A, m_Z$ and $\tan\beta$ as in
Eq.(\ref{relation of the mass parameters}).
\subsection{Appendix C}
Here we list some of the couplings of Higgs-quark-quark and
Higgs-squark-squark vertices which were used in the analysis of
the numerical results:
\begin{eqnarray}
\label{the relevant vertices}
A^0-\bar{t}-t:~~~~&& V_{A^0\bar{t}t}\gamma^5=-\frac{g m_t \cot\beta}{2 m_W} \gamma^5 \nb \\
H^0-\bar{t}-t:~~~~&& V_{H^0\bar{t}t}=-i \frac{g m_t \sin\alpha}{2 m_W \sin\beta} \nb \\
A^0-\tilde{t}_i^{\dag}-\tilde{t}_j:~~~~&&
V_{A^0\tilde{t}_i^{\dag}\tilde{t}_j}=
    -\frac{g m_t}{2 m_W}(\mu + A_t \cot\beta)
    ({\cal R}^{\tilde{t}}_{i1} {\cal R}^{\tilde{t}}_{j2}
    -{\cal R}^{\tilde{t}}_{i2} {\cal R}^{\tilde{t}}_{j1}) \nb \\
A^0-\tilde{b}_i^{\dag}-\tilde{b}_j:~~~~&&
V_{A^0\tilde{b}_i^{\dag}\tilde{b}_j}=
    -\frac{g m_b}{2 m_W}(\mu + A_b \tan\beta)
    ({\cal R}^{\tilde{b}}_{i1} {\cal R}^{\tilde{b}}_{j2}
    -{\cal R}^{\tilde{b}}_{i2} {\cal R}^{\tilde{b}}_{j1})
\end{eqnarray}
(1) For large $\tan\beta (\gtrsim 30)$, the coupling
$V_{A^0\bar{t}t}$ is suppressed due to the factor $\cot\beta$,
therefore the radiative corrections to the decays $A^0 \rightarrow
\tilde{\chi}_1^+ \tilde{\chi}_1^-/\tilde{\chi}_2^0
\tilde{\chi}_2^0$
contributed by the diagrams involving the vertex $A^0-\bar{t}-t$ are negligible. \\
(2) The couplings $V_{A^0\tilde{t}_i^{\dag}\tilde{t}_j}$ and
$V_{A^0\tilde{t}_i^{\dag}\tilde{t}_j}$ are zero when $i=j$.

\vskip 10mm
\begin{flushleft} {\bf Figure Captions} \end{flushleft}

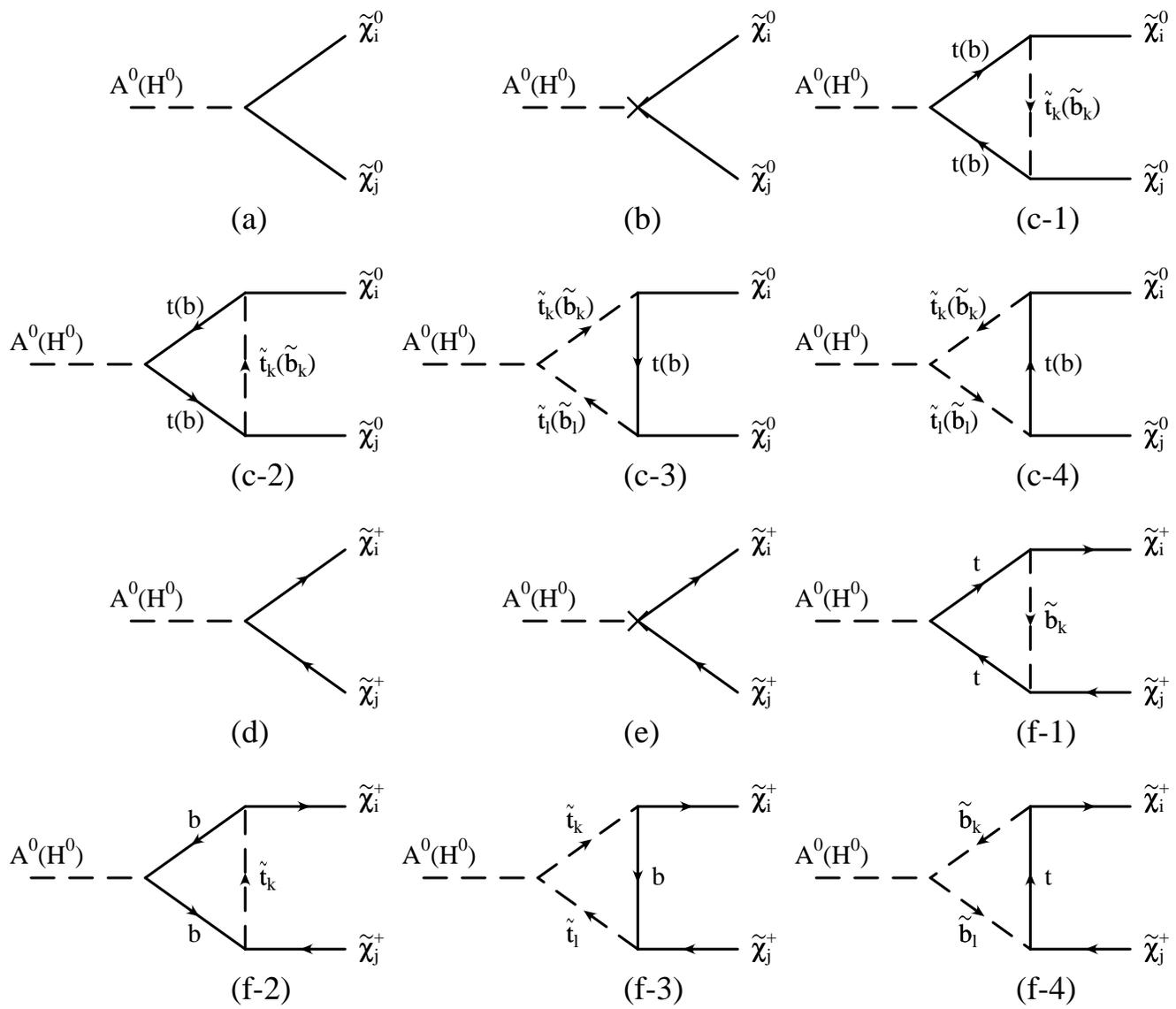
{\bf Fig.1} The relevant Feynman diagrams to the decays $A^0/H^0
\rightarrow \tilde{\chi}_i^0 \tilde{\chi}_j^0$, $A^0/H^0
\rightarrow \tilde{\chi}_i^+ \tilde{\chi}_j^-$: (a) and (d)
tree-level diagrams; (b) and (e) counterterms for vertices;
(c-1)-(c-4) and (f-1)-(f-4) one-loop vertex corrections. In
diagrams (a), (b) and (c), $i$ = $j$ = 2, (d), (e) and (f), $i$ =
$j$ =1. In all the loop diagrams, the subscripts $k$ and $l$ run
from 1 to 2.
\par
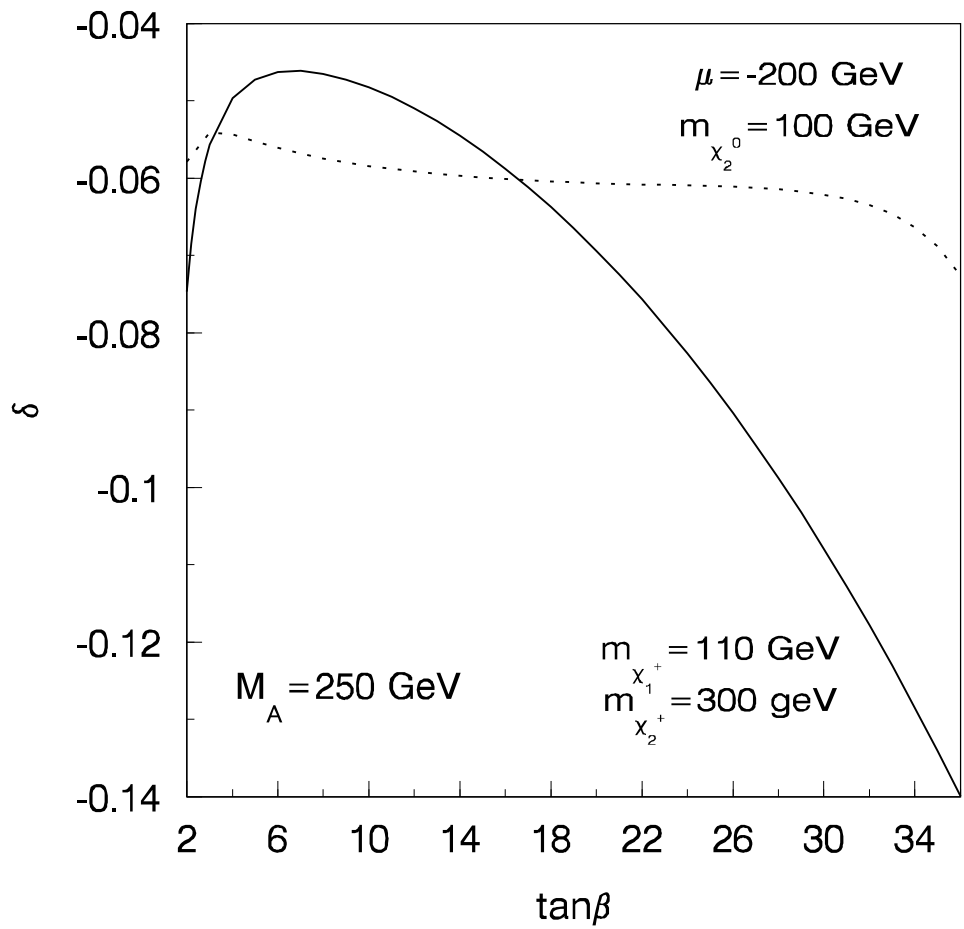
{\bf Fig.2} The relative correction to $A^0$ decay as a function
of $\tan\beta$.\ The solid and dotted curves represent the decays
$A^0 \rightarrow \tilde{\chi}_1^+ \tilde{\chi}_1^-$ and $A^0
\rightarrow \tilde{\chi}_2^0 \tilde{\chi}_2^0$ respectively.
\par
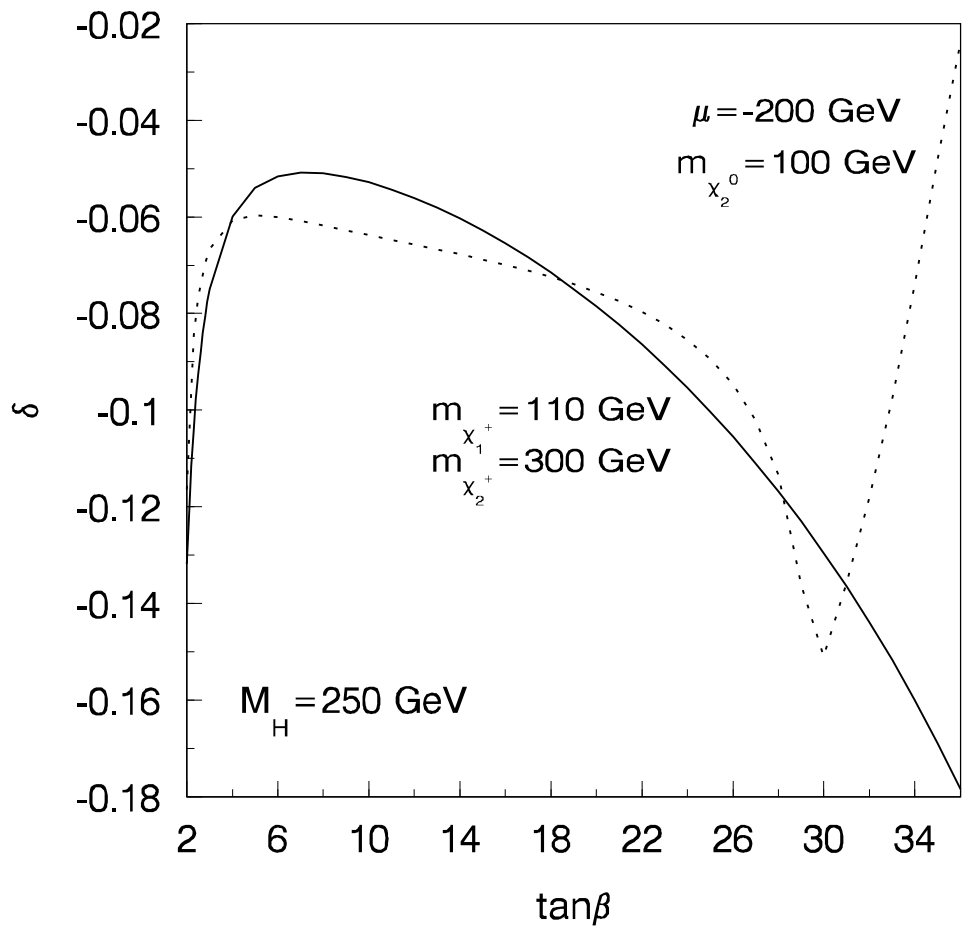
{\bf Fig.3} The relative correction to $H^0$ decay as a function
of $\tan\beta$. The solid and dotted curves represent the decays
$H^0 \rightarrow \tilde{\chi}_1^+ \tilde{\chi}_1^-$ and $H^0
\rightarrow \tilde{\chi}_2^0 \tilde{\chi}_2^0$ respectively.
\par
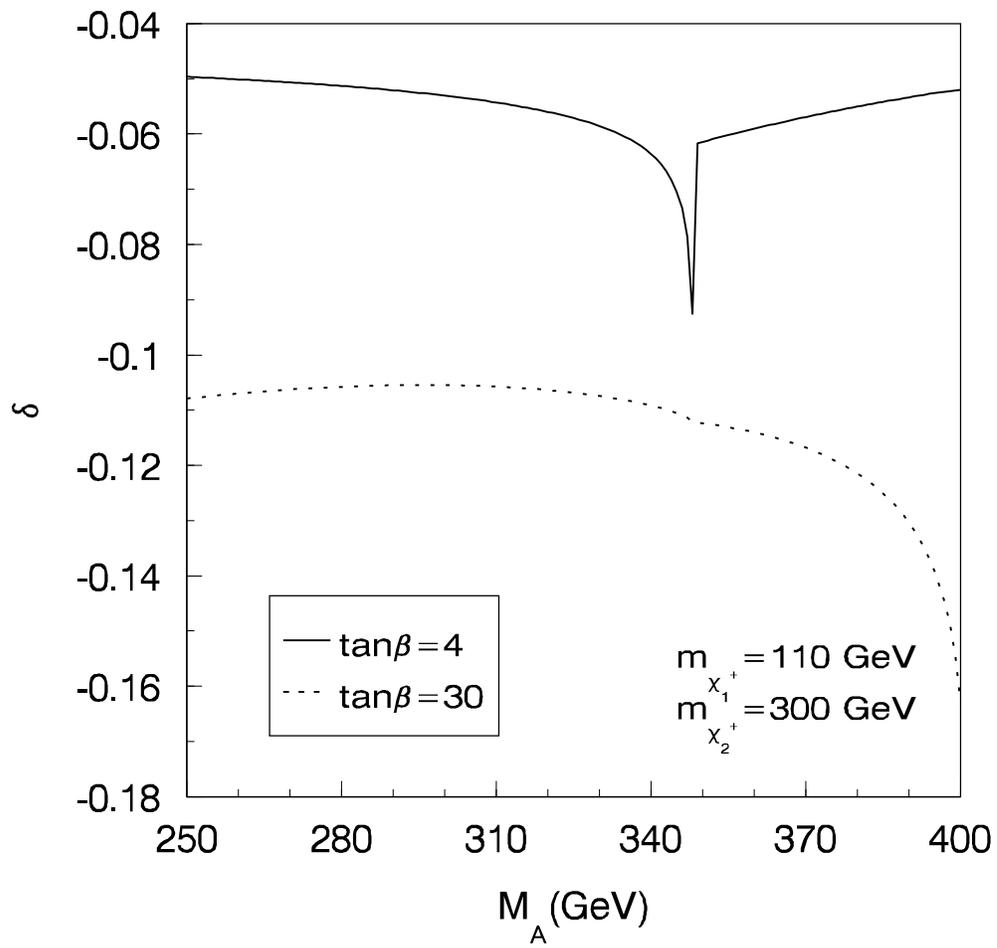
{\bf Fig.4} The relative correction to the decay $A^0 \rightarrow
\tilde{\chi}_1^+ \tilde{\chi}_1^-$ as a function of $m_A$.
\par
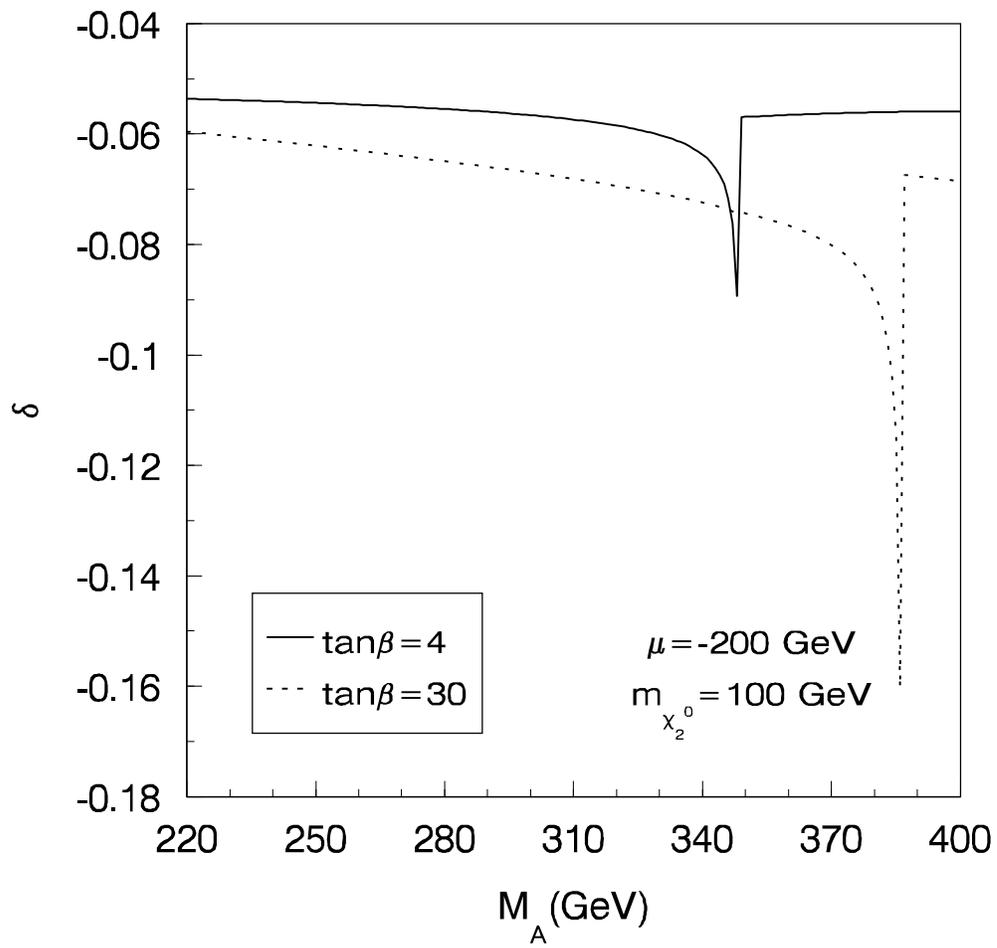
{\bf Fig.5} The relative correction to the decay $A^0 \rightarrow
\tilde{\chi}_2^0 \tilde{\chi}_2^0$ as a function of $m_A$.
\par
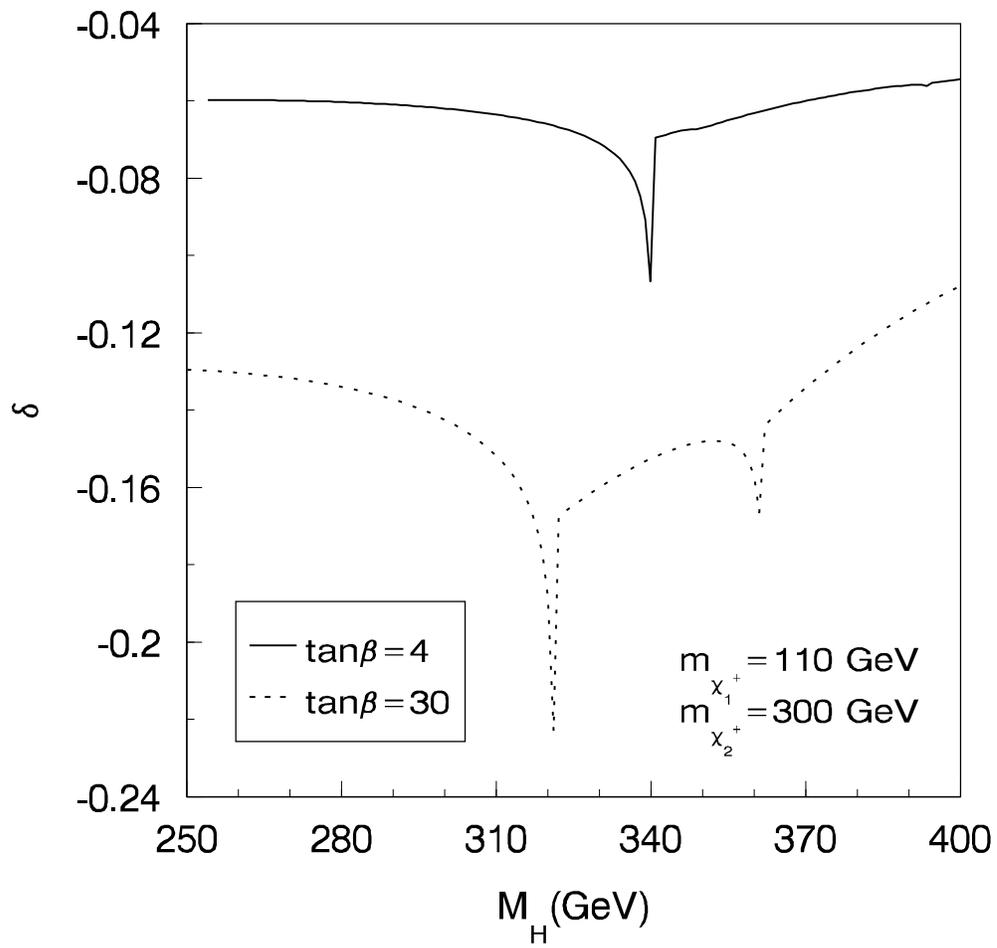
{\bf Fig.6} The relative correction to the decay $H^0 \rightarrow
\tilde{\chi}_1^+ \tilde{\chi}_1^-$ as a function of $m_H$.
\par
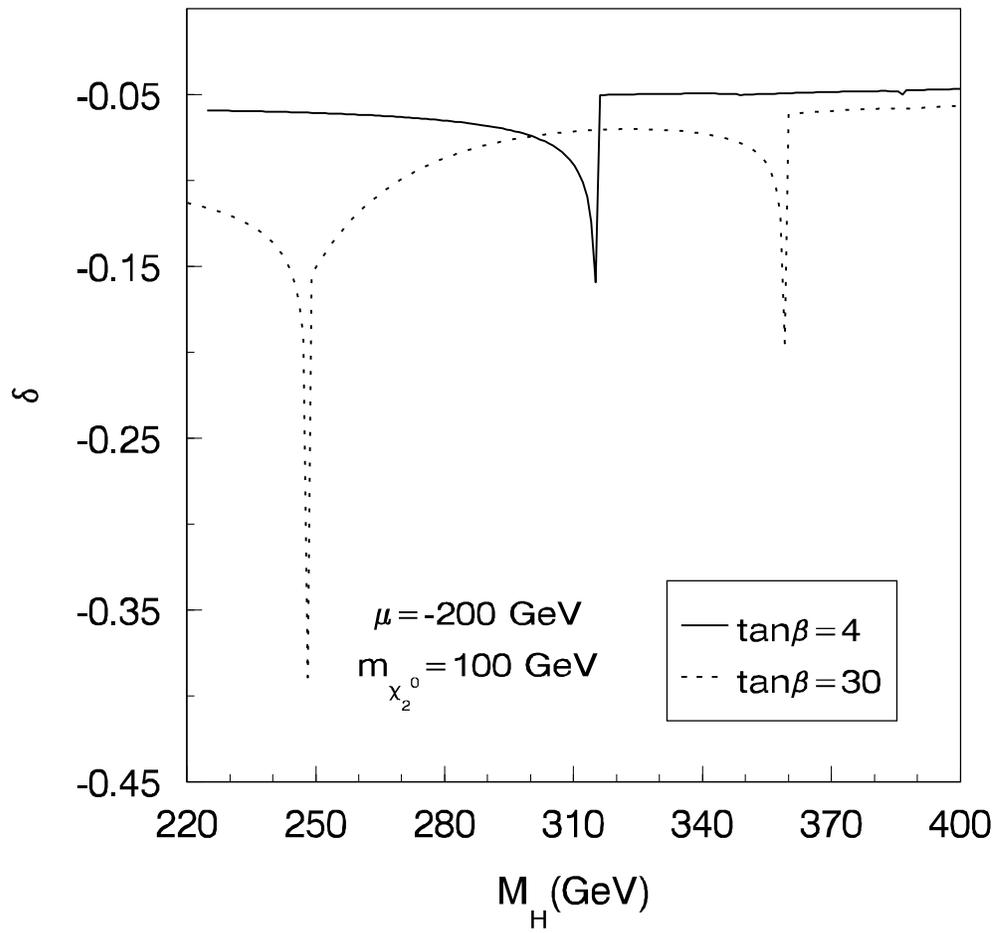
{\bf Fig.7} The relative correction to the decay $H^0 \rightarrow
\tilde{\chi}_2^0 \tilde{\chi}_2^0$ as a function of $m_H$.
\par
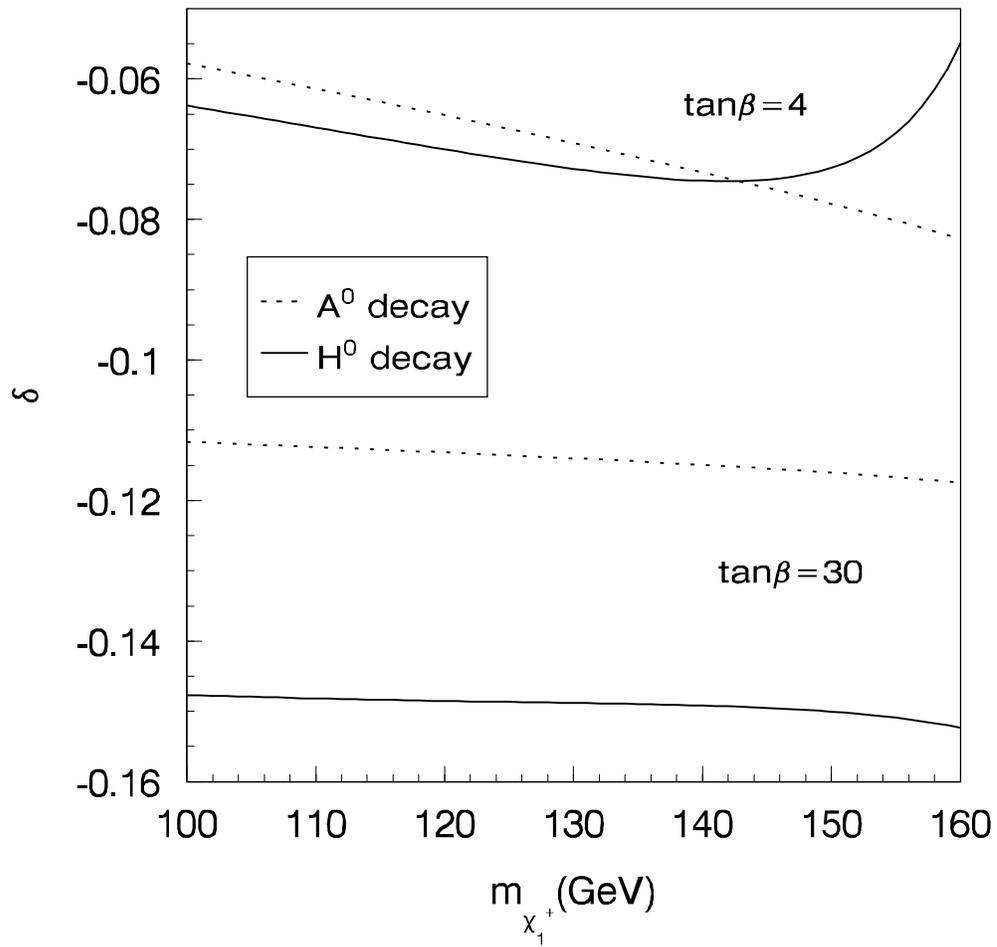
{\bf Fig.8} The relative correction as a function of
$m_{\tilde{\chi}_1^+}$.  The solid (dotted) curves represent the
decay $H^0 \rightarrow \tilde{\chi}_1^+ \tilde{\chi}_1^-$ ($A^0
\rightarrow \tilde{\chi}_1^+ \tilde{\chi}_1^-$) for $\tan\beta=4$
and 30. The input parameters are set to be $m_H = 350$ GeV ($m_A =
350$ GeV) and $m_{\tilde{\chi}_2^+} = 300$ GeV.
\par
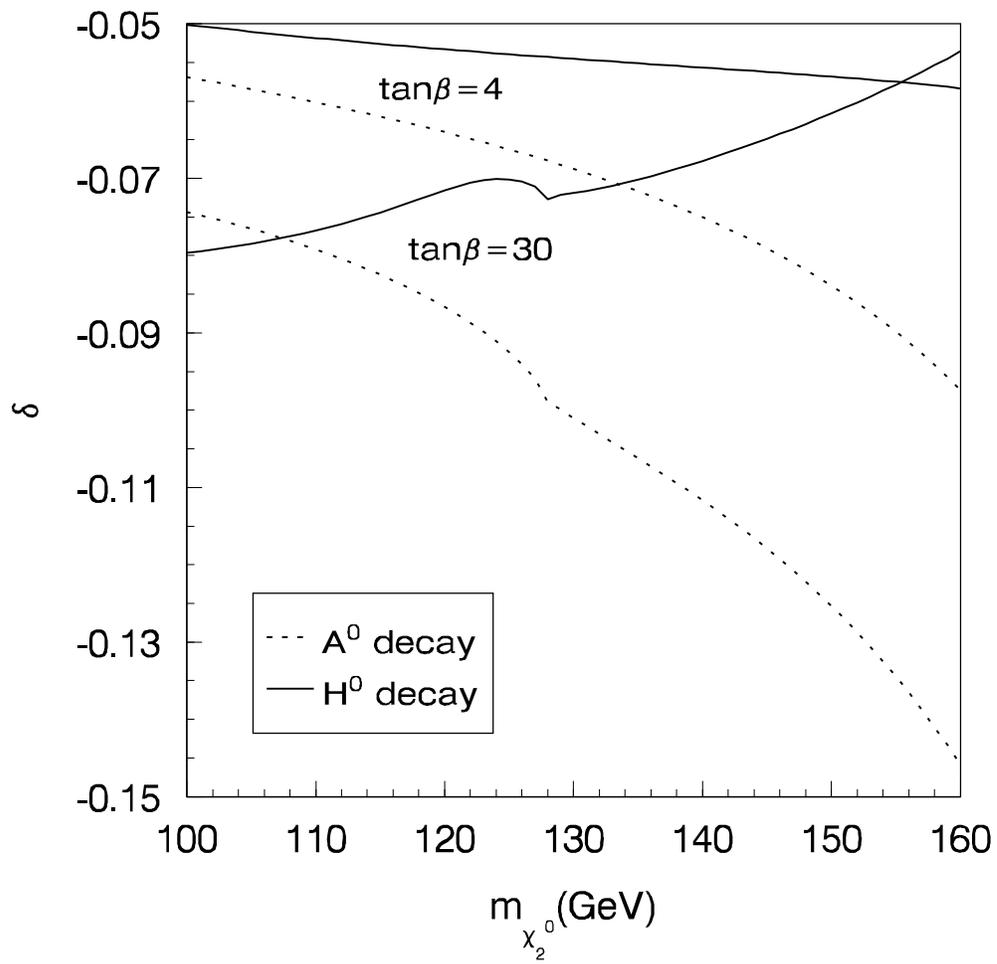
{\bf Fig.9} The relative correction as a function of
$m_{\tilde{\chi}_2^0}$.  The solid (dotted) curves represent the
decay $H^0 \rightarrow \tilde{\chi}_2^0 \tilde{\chi}_2^0$ ($A^0
\rightarrow \tilde{\chi}_2^0 \tilde{\chi}_2^0$) for $\tan\beta=4$
and 30. The input parameters are set to be $m_H = 350$ GeV ($m_A =
350$ GeV) and $\mu = -200$ GeV.
\end{document}